\documentclass[journal]{IEEEtran}

\usepackage{pifont,calc}
\usepackage{cite}

\usepackage{graphicx}
\ifCLASSINFOpdf
   \usepackage[pdftex]{graphicx}
\else
 \fi
\usepackage[cmex10]{amsmath}
\usepackage{array}

\usepackage{cases} 
\usepackage{bm} 

\makeatletter

\newcommand{\Rmnum}[1]{\expandafter\@slowromancap\romannumeral #1@}
\makeatother
\usepackage{booktabs}

\begin{document}

\title{A Fast Recursive Algorithm for G-STBC}

\author{Hufei Zhu, Wen Chen, Bin Li, and Feifei Gao,~\IEEEmembership{Members,~IEEE}
\thanks{H. Zhu and W. Chen are with Department of Electronic Engineering, Shanghai
Jiao Tong University, Shanghai 200240, P. R. China and Huawei
Technologies Co., Ltd., e-mail: zhuhufei@yahoo.com.cn and
wenchen@sjtu.edu.cn;}
\thanks{H. Zhu is also with SKL of ISN, Xidian University, and W.
Chen is also with SKL of Mobile Communications, Southeast
University, P. R. China;}
\thanks{B. Li is with Huawei Technologies Co.,
Ltd., Shenzhen 518129, P.R. China, e-mail: binli@huawei.com;}
\thanks{F. Gao is with the Department of Automation, Tsinghua University, China,
and is also with the School of Engineering and Science, Jacobs University Bremen,
Germany. Email: feifeigao@ieee.org.}
\thanks{This work is supported by NSF China \#60972031, by SEU SKL project
\#W200907, by ISN project \#ISN11-01, and by National 973 project
\#2009CB824904.}
}

\markboth{IEEE Transactions on Communications}%
{Shell \MakeLowercase{\textit{et al.}}: Bare Demo of IEEEtran.cls for Journals}

\maketitle

\begin{abstract}
This paper proposes a fast recursive algorithm for Group-wise
Space-Time Block Code
(G-STBC), which 
takes full advantage of the Alamouti structure in the equivalent
channel matrix to reduce the computational complexity. With respect
to the existing efficient algorithms for G-STBC,
the proposed
algorithm achieves better performance and
usually requires less computational complexity.
\end{abstract}


\IEEEpeerreviewmaketitle

\section{Introduction}
\IEEEPARstart{M}{ultiple}-input multiple-output (MIMO) wireless
communication systems possess huge channel capacities
\cite{Jun28_MIMO} when the multipath scattering is sufficiently
rich. A MIMO system can  provide two types of gains simultaneously,
i.e. spatial multiplexing gain and diversity gain, while
usually the increase in one type of gain needs the sacrifice in the
other type of gain \cite{DivstMltplxTdof}. Two extreme examples of
MIMO systems are Vertical Bell Laboratories Layered Space-Time
architecture (V-BLAST) \cite{zhf1} and space-time block code (STBC)
\cite{fAlamoutiSTBC,Jun09add_GeneralSTBC}, which achieve full
spatial multiplexing gain and full diversity gain, respectively. The
combination of V-BLAST and STBC is known as Group-wise STBC (G-STBC)
or Multi-layered STBC (MLSTBC) \cite{SMSTBC1,
SMSTBC2,SMSTBC3,SMSTBC4,SMSTBC5,SMSTBC6,Jun09add_bestCited14,Jun09add_bestCited15,Jun09add_TransCom},
which achieves both spatial multiplexing gain and diversity gain
concurrently. 
Specifically, the Alamouti code (the simplest and most popular STBC)
\cite{fAlamoutiSTBC} and the G-STBC consisting of two Alamouti codes
\cite{SMSTBC2,Jun09add_bestCited14,Jun09add_bestCited15,Jun09add_TransCom}
have been adopted
in wireless standards \cite{Jun09add_wimax},
where they are named space-time transmit diversity (STTD) and
double-STTD (DSTTD), respectively.


In G-STBC, the transmit antennas are divided into $M$ layers
(groups), of which each has $K$ transmit antennas and a
corresponding STBC encoder. To detect one layer by suppressing the
interferences from the other $M-1$ layers, the detector in
\cite{SMSTBC1} requires $N\ge(M-1)\times K+1$ receive antennas,
while the linear or successive interference cancelation (SIC)
detectors in
\cite{SMSTBC2,SMSTBC3,SMSTBC4,SMSTBC5,SMSTBC6,Jun09add_bestCited14,Jun09add_bestCited15,Jun09add_TransCom}
only require $N\ge M$ receive antennas, which 
are developed from the equivalent channel model based on the
temporal and spatial structure of STBC. Moreover, group-wise SIC
detectors and trellis-coded modulation (TCM) encoders/decoders are
jointly designed  for G-STBC
in \cite{WXD_revier1_ref}. The minimum mean-square error (MMSE)
ordered SIC (OSIC) detector for G-STBC \cite{SMSTBC3} and that only
for DSTTD \cite{SMSTBC2} both require high computational
complexities.
Then some efficient detectors for G-STBC or only for DSTTD are
proposed in
\cite{SMSTBC4,SMSTBC5,SMSTBC6,Jun09add_bestCited14,Jun09add_bestCited15,Jun09add_TransCom}.
The MMSE sub-optimal OSIC detector for V-BLAST \cite{QR_BLAST} using
sorted QR decomposition (SQRD) are extended to decode G-STBC in
\cite{SMSTBC4}. The simple linear Zero Forcing (ZF) detector for
G-STBC in \cite{SMSTBC5} is further simplified in \cite{SMSTBC6} by
utilizing Strassen algorithm.
For DSTTD, 
SIC detectors are reported in
\cite{Jun09add_bestCited14,Jun09add_bestCited15}.
Recently the authors of \cite{Jun09add_TransCom} notice that none of
the DSTTD detectors in
\cite{SMSTBC2,Jun09add_bestCited14,Jun09add_bestCited15} utilizes
the post-cancellation orthogonal structure, and then utilizes that
structure to develop the 
SIC detector \cite{Jun09add_TransCom} for DSTDD that reduces the
complexity at the sacrifice of performance.

In this letter, we consider the G-STBC consisting of $M\ge 2$
Alamouti Codes. We deduce an efficient recursive algorithm to
compute the initial estimation error covariance matrix, based on
which we propose a fast recursive algorithm for G-STBC. The proposed
algorithm 
exploits 
the Alamouti structure \cite{Invariant_Alamouti} in the equivalent
channel matrix to reduce the 
complexity dramatically.


In what follows,
$( \bullet )^T$, $ ( \bullet )^* $, and $ ( \bullet )^H$ 
denote matrix transposition, matrix conjugate, and matrix conjugate
transposition, respectively. $ {\bf{{\rm I}}}_M $ is the identity
matrix of size $M$.

\section{System Model}
The considered G-STBC system consists of $M\times K$ transmit
antennas and $N$ ($\ge M$) receive antennas in a rich-scattering and
flat-fading wireless channel.  It includes $M$ parallel and
independent STBC encoders. In this letter, we consider the Alamouti
STBC encoder \cite{fAlamoutiSTBC} with $K=2$. The transmitted data
stream
${\bf{s'}} = \left[ {\begin{array}{*{20}c}
   {s_{11} } & {s_{12} } & {s_{21} } & {s_{22} } &  \cdots  & {s_{M1} } & {s_{M2} }  \\
\end{array}} \right]^T$
is de-multiplexed into $M$ sub-streams. Each sub-stream
$s_{m1},s_{m2}$ ($m=1,2,...,M$) is encoded by an independent
Alamouti encoder, and then fed to its respective $K=2$ transmit
antennas.
Correspondingly the received symbols over two time slots are
\begin{multline}\label{rxGSTBC9Nv23} 
\left[ {\begin{array}{*{20}c}
   {x_{11} } & {x_{12} }  \\
   {x_{21} } & {x_{22} }  \\
    \vdots  &  \vdots   \\
   {x_{N1} } & {x_{N2} }  \\
\end{array}} \right] = \left[ {\begin{array}{*{20}c}
   {h_{11} } & {h_{12} } &  \cdots  & {h_{1,2M} }  \\
   {h_{21} } & {h_{22} } &  \cdots  & {h_{2,2M} }  \\
    \vdots  &  \vdots  &  \ddots  &  \vdots   \\
   {h_{N1} } & {h_{N2} } &  \cdots  & {h_{N,2M} }  \\
\end{array}} \right]\times
\\
\left[{\begin{array}{*{20}c}
   {s_{11}^{}} & {s_{12}^{}} &  \cdots &  {s_{M1}^{}} & {s_{M2}^{}}\\
   {-s_{12}^*} & {s_{11}^* } &  \cdots & {-s_{M2}^*} & {s_{M1}^*} \\
\end{array}}\right]^T + {\bf{N}},
\end{multline}
where $x_{nj}$ 
is the $j^{th}$ symbol received by the $n^{th}$ receive antenna,
$h_{nm}$ 
is the fading gain from the $m^{th}$ transmitter 
to the $n^{th}$ receiver, 
and ${\bf{N}}$ is the $N\times 2$
complex Gaussian noise matrix. 
We can transform (\ref{rxGSTBC9Nv23}) into the equivalent channel
 model \cite{SMSTBC2} 
 \begin{equation}\label{equ:5}
{\bf{x'}} = {\bf{H's'}} + {\bf{n'}},
\end{equation}
where the equivalent received symbol vector
${\bf{x'}} = \left[ {\begin{array}{*{20}c}
   {x_{11} } & {x_{12}^* } & {x_{21} } & {x_{22}^* } &  \cdots  & {x_{N1} } & {x_{N2}^* }  \\
\end{array}} \right]^T$,
and the equivalent channel matrix
\begin{eqnarray}\label{equ:7}
\footnotesize
{\bf{H'}}=\left[\setlength{\arraycolsep}{1.0mm}
\renewcommand{\arraystretch}{1.0}
{\begin{array}{*{20}c}
   {h_{11} } & {h_{12} } &  \cdots  & {h_{1,2M - 1} } & {h_{1,2M} }  \\
   {h_{12}^* } & { - h_{11}^* } &  \cdots  & {h_{1,2M}^* } & { - h_{1,2M - 1}^* }  \\
    \vdots  &  \vdots  &  \ddots  &  \vdots  &  \vdots   \\
   {h_{N1} } & {h_{N2} } &  \cdots  & {h_{N,2M - 1} } & {h_{N,2M} }  \\
   {h_{N2}^* } & { - h_{N1}^* } &  \cdots  & {h_{N,2M}^* } & { - h_{N,2M - 1}^* }  \\
\end{array}} \right].
\end{eqnarray}
In (\ref{equ:5}), the transmitted data stream ${\bf s'}$ is with the covariance $E({\bf{s's'}}^H ) = \sigma _{{s}}^2 {\bf{{\rm I}}}_{2M}$,
and the $2N\times 1$ Gaussian noise vector ${\bf{n'}}$ is with zero
mean and the covariance $E({\bf{n'n'}}^H)=\sigma _n ^2 {\bf{{\rm I}}}_{2N}$.

The MMSE OSIC detector for V-BLAST can be extended to the equivalent
channel model (\ref{equ:5}) for G-STBC, since mathematically
(\ref{equ:5}) has the same form as the standard channel model for
V-BLAST \cite{SMSTBC3}.
Correspondingly the linear MMSE estimate of $\bf{s'}$ can be  \cite{SMSTBC3,SMSTBC4} 
 \begin{equation}\label{equ:2}
{\bf{y}} = \left( {{\bf{H'}}^H {\bf{H'}} + \alpha {\bf{I}}_{2M}}
\right)^{ - 1} {\bf{H'}}^H {\bf{x'}}={\bf{Q'}}{\bf{H'}}^H{\bf{x'}},
\end{equation}
where $\alpha  = \sigma _n^2 /\sigma _s^2$, and
 ${\bf{Q'}}=\left({{\bf{H'}}^H {\bf{H'}} + \alpha {\bf{I}}_{2M}}\right)^{ -
 1}$
is the estimation error covariance matrix.

The 
OSIC detector for G-STBC \cite{SMSTBC3} detects $2M$ entries of
$\bf{s'}$ iteratively with the optimal order.
In each
iteration, the entry with the highest 
SINR (Signal to Interference plus Noise Ratio)
is detected, and then its
interference is cancelled in
${\bf{x'}}$.
Suppose that
the entries of
${\bf{s'}}$
are
permuted
such that
the successive detection order is ${s_{M2}}, {s_{M1}}, {s_{(M-1)2}}, {s_{(M-1)1}}, \cdots, {s_{12}}, {s_{11}}$. Then in the $(2M-k+1)^{th}$ ($k=2M,2M-1,\cdots,1$) iteration, the $k$  undetected entries can be represented as ${\bf{s'}}_k$ that includes the first $k$ entries of ${\bf{s'}}$, while the corresponding equivalent channel matrix
 \begin{equation}\label{equ:def_H_k_col}
{\bf{H'}}_k = \left[{\begin{array}{*{20}c}
   {{\bf{h'}}_1 } & {{\bf{h'}}_2 } &  \cdots  & {{\bf{h'}}_{k}} \\
\end{array}} \right],
\end{equation}
where ${\bf{h}}'_i$ ($i=1,2,\cdots,k$) denotes the $i^{th}$ column
of ${\bf{H'}}$.
%


\section{A Fast Recursive Algorithm for G-STBC}
In this section, we deduce an efficient recursive algorithm to
compute the initial ${\bf{Q'}}$, based on which we develop a fast
recursive algorithm for G-STBC. 

The $2\times 2$ Alamouti structure
$\left[\setlength{\arraycolsep}{0.5mm}
\renewcommand{\arraystretch}{0.5}
{\begin{array}{*{20}c}
   {a_1 } & { - a_2^* }  \\
   {a_2 } & {a_1^* }  \\
\end{array}} \right]$ in
 the equivalent channel matrix
${{\bf{H'}}}$,
which remains invariant under several matrix operations
\cite{Invariant_Alamouti}, 
is regarded as the matrix
representation of a quaternion and then called as a block entry
\cite{Invariant_Alamouti}. Block entries can form vectors and
matrices, which are then called as block vectors and block matrices
\cite{Invariant_Alamouti}, respectively. In what follows, 
scalars, vectors and matrices with overlines
always represent the above-mentioned block entries (i.e. block
scalars), block vectors, and block matrices, respectively.
 We 
 define the block matrix
\begin{equation}\label{Jan9def_BarH_m}
{\bf{\bar {H}  }}_m={\bf{H'}}_{2m}=\left[ {\begin{array}{*{20}c}
   {{\bf{\bar h}}_1 } & {{\bf{\bar h}}_2 } &  \cdots  & {{\bf{\bar h}}_m }  \\
\end{array}} \right],
\end{equation}
 where $m=1,2,\cdots,M$, and the block column vector ${\bf{\bar h}}_m  = \left[{\begin{array}{*{20}c}
   {{\bf{h'}}_{2m - 1} } & {{\bf{h'}}_{2m} }  \\
\end{array}}\right]$.
Then we write
\begin{subnumcases}{\label{Oct21equ19a22}}
{\bf{{\bar R}}}_{m} = {{\bf{\bar {H}}}_m}^{H} {\bf{\bar {H}}}_{m} + \alpha
{\bf{I}}_{2m} = \left[ {\begin{array}{*{20}c}
   {{\bf{{\bar R}  }}_{{m-1}}^{} } & {{\bf{ {\bar{v}}}}_{{m-1} } }  \\
   {{\bf{{\bar{v}}}}_{{m-1} }^H } & {{{{\bar {{\upsilon}}} }} _{m}}  \\
\end{array}} \right], &  \label{equ:19}\\
{\bf{{{\bar Q}  }}}_{m} = {\bf{{\bar R} }}_{m}^{
- 1} = \left[ {\begin{array}{*{20}c}
   {{\bf{{{\bar T}  }}}_{m-1}^{} } & {{{\bf{\bar {w}}}}_{m-1} }  \\
   {{{\bf{\bar {w}}}}_{m-1}^H } & {{{{\bar {{\omega}}}}} _{m} }  \\
\end{array}} \right], & \label{equ:22}
\end{subnumcases}
where ${\bf{\bar R}}_{m-1}$ and ${\bf{\bar T}}_{m-1}$ are the
 leading principal $2(m-1)\times 2(m-1)$ sub-matrices
 of ${\bf{\bar R}}_{M}$ and ${\bf{\bar Q}}_{m}$ \cite{My_IC_In_z, zhf6}, respectively.
 In (\ref{equ:19}),
  ${\bf{\bar {H}  }}_m^H {{\bf{\bar {H}  }}_m}$ (computed from the block matrix ${{\bf{\bar {H}
  }}_m}$)
  must be Hermitian with diagonal $2\times 2$ sub-blocks that are non-negative scaled multiples
  of the identity matrix and with off-diagonal $2\times 2$ sub-blocks that are  Alamouti
  matrices \cite[Property 3 in the subsection ``B. Block Matrices"]{Invariant_Alamouti}, and
  the same is ${\bf{{\bar R}  }}_{m}$. Then in (\ref{equ:19}), the block entry
\begin{equation}\label{equ:20}
{{{\bar {{\upsilon}}}}}_{{m}}={{{\upsilon}'}}  _{m} {\bf{I}}_2
 \end{equation}
  (where ${{{\upsilon}'}}_{m}$ is a scalar),
  while ${\bf{{\bar{v}}}}_{{m-1}}$ is the block column
  vector, which can be represented as
${\bf{{\bar{v}}}}_{{m-1} }  = \left[ {\begin{array}{*{20}c}
   {{\bf{v'}}_{2(m-1)} } & {{\bf{v''}}_{2(m-1)} }  \\
\end{array}} \right].$
Moreover, in Appendix A we derive ${{{\bar {{\omega}}} }}_{m}$,
${{\bf{\bar {w}}}}_{m-1}$ and ${\bf{{{\bar T}  }}}_{m-1}^{}$ in
(\ref{equ:22}) respectively as
\begin{subnumcases}{\label{derive5}}
{{\omega}'} _{m} = \left( {{{{\upsilon}'}} _{m}  -
{\bf{{v'}}}_{2(m-1)}^H {\bf{{{\bar Q}  }}}_{m-1}
{\bf{{v'}}}_{2(m-1)} } \right)^{ -
1}, & \label{derive5more}\\
{{{\bar {{\omega}}} }}_{m}  = {{\omega}'} _{m} {\bf{I}}_2, & \label{derive5a}\\
{{\bf{\bar {w}}}}_{m-1} =  - {{\omega}'} _{m} {\bf{{{\bar Q}
}}}_{m-1}{\bf{{\bar{v}}}}_{m-1}, &  \label{derive5b}\\
{\bf{{{\bar T}  }}}_{m-1}^{}  = {\bf{{{\bar Q}  }}}_{m-1}^{}  +
{{\omega}'} _{m}^{ - 1} {{\bf{\bar {w}}}}_{m-1}^{} {{\bf{\bar
{w}}}}_{m-1}^H. & \label{derive5c}
\end{subnumcases}

Base on (\ref{derive5}), we develop the recursive G-STBC algorithm.
In the \textsl{initialization} phase, to obtain ${\bf{{{\bar Q}
}}}_M$ from ${\bf{{{\bar Q} }}}_{1}={\bf{{{\bar R} }}}_{1}^{-1}$, we
compute ${\bf{{{\bar Q} }}}_m$ from ${\bf{{{\bar Q}  }}}_{m-1}$
recursively (for $m=2, 3,..., M$) by (\ref{derive5})  and
(\ref{equ:22}).
Moreover, as the V-BLAST algorithms in \cite{My_IC_In_z,zhf6}, we
perform interference cancellation not in ${\bf{x'}}$, but in ${\bf
z}'={\bf{\bar{H}}}_M^H {\bf{x'}}$. Then we need to compute the
initial
\begin{equation}\label{Oct22DefzMstbc}
{\bf z}'_M={\bf{\bar{H}}}_M^H {\bf{x'}}.
\end{equation}

In the \textsl{recursion} phase,
we detect $M$ layers of STBC encoded symbols recursively by
group-wise  OSIC.
The subscript $|$ is added
to some variables for the \textsl{recursion} phase, to distinguish
them from the variables for the \textsl{initialization} phase,
e.g., usually ${\bf{\bar Q}}_{|m}\ne {\bf{\bar Q}}_{m}$. In each recursion, estimate the $p_{m}^{th}$
layer,
i.e. the layer with the highest SINR among all the undetected
layers, by
\begin{equation}\label{equ:27}
\left[ {\begin{array}{*{20}c}
   {y_{p_{m} 1}} & {y_{p_{m} 2}}  \\
\end{array}} \right]^T  = \left[ {\begin{array}{*{20}c}
   {{\bf{{q}}}'_{|2{ m}-1}} & {{\bf{{q}}}'_{|2{ m}}}  \\
\end{array}} \right]^H {\bf z}'_{ m},
\end{equation}
where  ${\bf{{q}}}'_{|2{m}-1}$ and ${\bf{{q}}}'_{|2{m}}$
  are the $\left({2m - 1}\right)^{th} $  and $ 2m^{th} $
  columns of  the permuted ${\bf{{{\bar Q}  }}}_{|m}$, respectively.
  Quantize ${y_{p_{m} 1} }$ and ${y_{p_{m} 2}}$ to obtain
  ${\hat s_{p_{m} 1} }={Q\left\{ {y_{p_{m} 1} } \right\}}$ and ${\hat s_{p_{m} 2}}={Q\left\{ {y_{p_{m} 2} } \right\}}$, respectively.
Then as in \cite{My_IC_In_z,zhf6}, we cancel the effect of
${s_{p_{m} 1} }$ and ${s_{p_{m} 2} }$ in ${\bf z}'_{ m}$ by
\begin{equation}\label{equ:STBC_IC}
{\bf z}'_{{ m}-1}={{\bf z'}_{ m}^{\left[ { - 2}
\right]}}
  - {\bf{{\bar{v}}}}_{{| m-1}}\left[ {\begin{array}{*{20}c}
   {\hat s_{p_{m} 1}} & {\hat s_{p_{m} 2}}  \\
\end{array}} \right]^T,
 \end{equation}
where  ${\bf z'}_{m}^{\left[ { - 2} \right]}$ is $ {\bf z}'_{ m}$
with
the last two entries removed, and 
${\bf{{\bar{v}}}}_{{|m-1}}$ is in ${\bf{{\bar R} }}_{| m}$, as shown
in (\ref{equ:19}). In the next recursion,
${\bf{{{\bar Q}  }}}_{| m-1}$ is required to compute (\ref{equ:27}).
Thus we deflate ${\bf{{{\bar Q}}}}_{|m}$ by (\ref{derive5c}) to
obtain
\begin{equation}\label{deflat_STBC_Q}
{\bf{{{\bar Q}  }}}_{| m-1}  = {\bf{{{\bar T}  }}}_{| m-1} -
{{\omega}'} _{| m}^{ - 1} {{\bf{\bar {w}}}}_{| m-1}^{} {{\bf{\bar
{w}}}}_{|m-1}^H.
 \end{equation}

To some extent, (\ref{deflat_STBC_Q}) (that deflates a block matrix
recursively) has a similar form as equation
(15) in \cite{TaiwanPaper} and equation (21) in \cite{zhf6} (that deflates a matrix recursively), 
while the vector ${{\bf{{w}}}}_{m-1}$ in the equations of
\cite{TaiwanPaper,zhf6} is replaced with the block vector
${{\bf{\bar {w}}}}_{|m-1}^{}$ in (\ref{deflat_STBC_Q}). On the other
hand, as (\ref{deflat_STBC_Q}), equation (24) in
\cite{Reviewer2_Journ_real_Gstbc} also deflates a block matrix
recursively, while it deflates the matched-filtered (MF) real-valued
channel matrix for G-STBC that is obtained from the augmented
real-valued channel matrix.

Table \Rmnum{1} summarizes the proposed recursive G-STBC algorithm,
where ${r'}_{|i,k}^{m}$ and ${q'}^{m}_{|i,k}$ are the entries in the
$i^{th}$ row and $k^{th}$ column of the permuted ${\bf{\bar R}}_{|
m}$ and ${\bf{\bar Q}}_{| m}$, respectively. Notice that
${\bf{{{\bar Q} }}}_{m}$($={\bf{{{\bar R} }}}_{m}^{-1}$) and
${\bf{{{\bar Q}  }}}_{| m}$
consist
of Alamouti sub-blocks\cite[Lemma ``Invariance Under
Inversion"]{Invariant_Alamouti}.

\begin{table*} [!t]
\renewcommand{\arraystretch}{1.3}
\caption{The proposed recursive G-STBC algorithm}
\label{table_IIoct22} \centering
\begin{tabular}{r l}
\hline \bfseries  Initialization   &Compute ${\bf z}'_M$ by
(\ref{Oct22DefzMstbc}). Compute the initial ${\bf{\bar
R}}_M={\bf{\bar R}}_{|M}$.
 Compute ${\bf{\bar Q}}_{m}$ from ${\bf{\bar Q}}_{m-1}$ recursively by (\ref{derive5})   \\
 \bfseries    & and (\ref{equ:22}) for $m=2,3, \cdots,M$, to obtain the initial ${\bf{\bar Q}}_M={\bf{\bar Q}}_{|M}$.   \\
\hline \bfseries Recursion  &Set ${\bf{p}} = \left[1, 2,\cdots,
M\right]^T$, and let $p_{m}$ denote the $m^{th}$ entry of
${\bf{p}}$.
For ${m}=M,M-1,\cdots,2$:  \\
\bfseries    &(a) Find $l_{m}={\arg \min}_{i = 2,4,\cdots}^{2{m}}
({q'}^{m}_{|i,i})$, since the entry with the highest SINR is the one
with the least \\
\bfseries &\qquad mean-square error \cite{My_IC_In_z,zhf6}. In
${\bf{{{\bar Q}}}}_{|m}$ and ${\bf{{{\bar R}
}}}_{|m}$, 
permute
 rows and columns ($l_{m}-1$, $l_{m}$)
with rows  \\
\bfseries     &\qquad and columns ($2m-1$, $2m$). 
Permute entries ($l_{m}-1$, $l_{m}$) with entries ($2m-1$, $2m$) in
${\bf z}'_{{m}-1}$. \\
\bfseries     &\qquad Interchange entries $l_{m}/2$ and ${m}$ of
${\bf{p}}$. \\
\bfseries     &(b) Compute ${y_{p_{m} 1}}$ and ${y_{p_{m} 2}}$ by
(\ref{equ:27}), which are quantized into
${\hat s_{p_{m} 1} }={Q\left\{ {y_{p_{m} 1} } \right\}}$ and ${\hat s_{p_{m} 2}}={Q\left\{ {y_{p_{m} 2} } \right\}}$. \\
\bfseries     &(c) Cancel the effect of $[{s_{p_{m} 1}}, {s_{p_{m}
2} }]$ in ${\bf z}'_{ m}$ by (\ref{equ:STBC_IC}), to obtain ${\bf
z}'_{{m}-1}$. \\
\bfseries     &(d) Deflate ${\bf{\bar Q}}_{| m}$ to get ${\bf{\bar
Q}}_{| m - 1}$ by (\ref{deflat_STBC_Q}). Remove the last two columns
and rows of
${\bf{\bar R}}_{| m}$ to get ${\bf{\bar R}}_{| m - 1}$. \\
\hline \bfseries  Solution  & When $m=1$, only execute the
above-described step (b).
The hard decisions of $[{s_{p_{m} 1}}, {s_{p_{m} 2}}]$ are \\
 & $[{\hat s_{p_{m} 1}}, {\hat s_{p_{m} 2}}]$, for ${m}=1,2,\cdots,M$. \\
\hline
\end{tabular}
\end{table*}

\begin{table*} [!t]
\renewcommand{\arraystretch}{1.3}
\caption{Complexities of the Presented G-STBC and DSTTD Algorithms}
\label{table_example} \centering
\begin{tabular}{c|c}
\bfseries  Algorithm (Alg.)     & \bfseries Total Complexity   \\
\toprule[1.2pt] \bfseries  Proposed G-STBC Alg.  &
${8M^{2}N+\frac{32}{3}M^{3}+O(M^2+MN)}$ real mult.,
\\
\bfseries  &  ${8M^{2}N+\frac{32}{3}M^{3}+O(M^2+MN)}$ real add. \\
\hline
\bfseries  G-STBC Alg. in \cite{SMSTBC4} &
 ${32M^3+16^2N+O(M^2+MN)}$ real
mult., \\
\bfseries  & ${32M^3+16M^2N+O(M^2+MN)}$ real add.
\cite[equation (19)]{SMSTBC4} \\
\midrule[1.2pt]
\bfseries  Proposed G-STBC Alg. with $M=N=3$ &  $570$ real mult.\\
\hline
\bfseries  ZF G-STBC Alg. in \cite{SMSTBC6} with $M=N=3$ &  $588$ real mult. \cite[Table \Rmnum{1}]{SMSTBC6} \\
\midrule[1.2pt]
\bfseries  Proposed DSTTD Alg. &  
$8M^{2}N+8MN+8N+67$ real mult.,
\\
\bfseries  &  $8M^{2}N+8MN+8N+40$ real add. ($M$ is always $2$ for DSTTD) \\
\hline
\bfseries  One-step SIC Alg. for   &  $\frac{8}{3}N^3+14N^2+\frac{79}{3}N-25$ real mult., \\
\bfseries  DSTTD in \cite{Jun09add_TransCom} &  $\frac{8}{3}N^3+10N^2+\frac{46}{3}N-9$ real add.\cite{Our_comments} \\
\bottomrule[1.2pt]
\end{tabular}\\
\end{table*}

\section{Performance Analysis and Numerical Results}
In the recursion phase,
the proposed group-wise OSIC
only needs to compute the block matrices ${\bf{\bar Q}}_{|m-1}$s
($m=M,M-1,\cdots,2$) that consist of Alamouti sub-blocks, while the
symbol-wise OSIC \cite{SMSTBC3} implemented by the recursive V-BLAST
algorithm in \cite{zhf6} needs to compute the matrices
${\bf{Q}}_{|k-1}$s ($k=2M,2M-1,2M-2,2M-3,\cdots,2$) that usually do
not consist of Alamouti sub-blocks. So 
the proposed group-wise OSIC can save much computational complexity,
with respect to the symbol-wise OSIC \cite{SMSTBC3}. On the other
hand, the proposed group-wise OSIC
is equivalent to the symbol-wise SIC
with the group-wise optimal detection order, as shown in Appendix B.
Thus its performance loss with respect to the symbol-wise OSIC (for
G-STBC) 
\cite{SMSTBC3}
only comes from different detection orders (i.e., the group-wise
optimal order vs. the symbol-wise
optimal order), and 
is relatively small, as shown in the rest of this section.



We list the computational complexities of the presented G-STBC and
DSTTD algorithms
in Table \Rmnum{2}, where
the exact complexities  of the presented DSTTD
algorithms 
have been verified by our numerical experiments to count the
floating-point operations (flops --- One flop can be one real
multiplication or one real addition). 
 In this table, ``Proposed DSTTD Alg." denotes the
proposed G-STBC algorithm applied to D-STTD. When counting the
complexities, we utilize the fact that one complex multiplication
needs $4$ real multiplications and $2$ real additions, while one
complex addition needs $2$ real additions. The total complexity of
the proposed G-STBC algorithm is the sum of the complexities to
compute ${\bf{\bar R}}_M$, ${\bf{\bar Q}}_M$ and ${\bf{\bar Q}}_{|
m-1}$s ($m=M,M-1,\cdots,2$) that are $\langle {2M^{2}N} \rangle$,
$\langle {2M^{3}} \rangle$ and $\langle {\frac{2}{3}M^{3}}\rangle$,
respectively, where $\langle {k} \rangle$ denotes $k$ complex
multiplications and $k$ complex additions. The algorithm in
\cite{SMSTBC4} nearly requires the same number of complex
multiplications and additions \cite{SMSTBC4}, \cite[Table
\Rmnum{1}]{QR_BLAST}, while \cite[equation (19)]{SMSTBC4} claims a
complexity of $2N_t^3 + 2N_t^2 N_r+O(N_t^2+N_t N_r)$ operations,
where $N_t=2M$, $N_r=N$, and an operation can be a complex
multiplication or addition. The complexity of the one-step SIC
algorithm for DSTTD in \cite{Jun09add_TransCom} has been revised in
\cite{Our_comments}, while the two-step SIC algorithm and the
SINR-ordered SIC algorithm for DSTTD in \cite{Jun09add_TransCom}
both need nearly double the complexity of the one-step SIC
algorithm.

For DSTTD, the special case of G-STBC with the minimum number of
transmit antennas, the ML(maximum likelihood)-like sphere detectors
\cite{Reviewer2_ML_receiver} are alternative methods worth
considering. For example, with respect to the SIC DSTTD detector
in~\cite{SMSTBC2}, the ML-like DSTTD detector in
\cite{Reviewer2_ML_receiver} performs about $1.7$ dB better at
BER=$10^{-2}$, but requires about $2.7$ times of complexity
\cite[Fig. 5]{Reviewer2_ML_receiver}. Although ML-like DSTTD
detectors can be preferred in some applications, they usually do not
have advantage in both complexity and BER performance
\cite{Reviewer2_ML_receiver}. Thus quite a few recent literatures
focusing on SIC DSTTD receivers
\cite{Jun09add_bestCited14,Jun09add_bestCited15,Jun09add_TransCom,WXD_revier1_ref}
did not discuss ML-like sphere detectors, and neither does this
letter focusing on SIC G-STBC receivers, which, due to the limited
space, cannot give 
too extensive discussion for only the special case of G-STBC (i.e.
DSTTD).


From Table \Rmnum{2}, we can compare the complexities of the
presented algorithms.
When $M = N$, the proposed G-STBC algorithm speeds up the G-STBC
algorithm in \cite{SMSTBC4} by $48/(\frac{56}{3})=2.57$
approximately, and even requires less real multiplications than the
linear ZF G-STBC algorithm in \cite{SMSTBC6}. When $N\ge 3$, the
proposed DSTTD algorithm is faster than the one-step SIC DSTTD
algorithm in \cite{Jun09add_TransCom}, and the speedup (in the
number of flops) grows with $N$, which is
 $1.02$ for $N=3$, $1.54$ for $N=4$, and $4.55$ for $N=8$. However, when $N=2$, the one-step SIC DSTTD
 algorithm in \cite{Jun09add_TransCom} (with much performance loss)
 is faster than the proposed DSTTD algorithm, while the speedup 
 is $1.76$. %

Let $N=M$. For different $N$, we carried out numerical experiments
to count the average flops per
time slot of 
the G-STBC algorithm in \cite{SMSTBC4} and the proposed G-STBC
algorithm
in Fig. 1. It can be seen that they are consistent with the
theoretical flops calculation.
%
On the other hand, Fig. 2 shows the BER (bit error rate) performance
of the algorithms in \cite{SMSTBC3,SMSTBC4}
and the proposed algorithm in a G-STBC system with $8$ 
transmit 
and $4$ receive antennas, while Fig. 3 and Fig. 4 show the BER
performance of the algorithms in
\cite{SMSTBC3,SMSTBC2,Jun09add_TransCom}
and the proposed algorithm in a DSTTD system with $2$ and $8$
receive antennas, respectively. We used QPSK modulation in Fig. 2,
 Fig. 3 and Fig. 4.
As shown in Fig. 2 and Fig. 3,
 the proposed
algorithm performs better than the efficient G-STBC algorithms in
\cite{SMSTBC4} and the efficient DSTTD algorithms in
\cite{Jun09add_TransCom}. %
For example, it performs about $0.4$ dB better than the SQRD
algorithm in \cite{SMSTBC4} at BER  $=10^{-4}$, and performs about $2$ dB better than the one-step 
fixed-order SIC algorithm 
for DSTTD in \cite{Jun09add_TransCom} at BER$=10^{-3}$. Fig. 2 and
Fig. 3 also shows that with respect to the MMSE OSIC algorithm in
\cite{SMSTBC3}, the proposed algorithm performs about $0.3$ dB worse
in the G-STBC system and $0.4$ dB worse in the DSTTD system, at BER
$=10^{-3}$. Moreover, it can be seen from Fig. 4 that for a large
number of receive antennas, the performance difference among the
presented DSTTD algorithms is small and even negligible.

\begin{figure}[!t]
\centering
\includegraphics[width=3.5in]{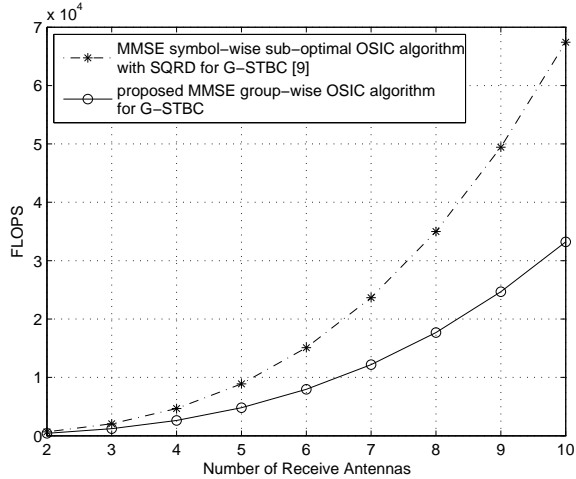}
\caption{Complexity Comparison between the G-STBC algorithm in
\cite{SMSTBC4} and the proposed G-STBC algorithm.} \label{fig flops}
\end{figure}

\begin{figure}[!t]
\centering
\includegraphics[width=3.5in]{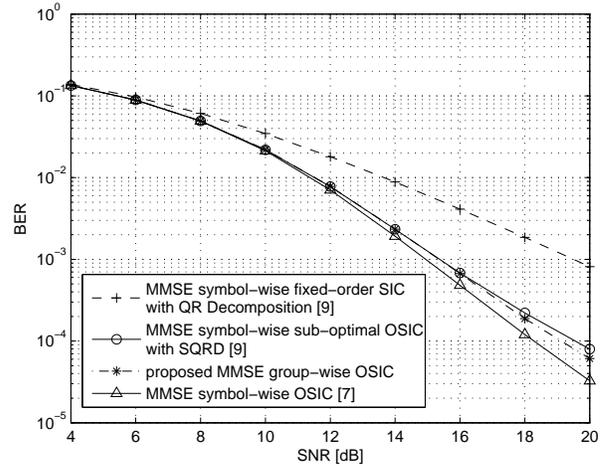}
\caption{Bit Error Rate (BER) comparison for a G-STBC system with
QPSK modulation, $8$ transmit and $4$ receive antennas.} \label{fig
flops}
\end{figure}

\begin{figure}[!t]
\centering
\includegraphics[width=3.5in]{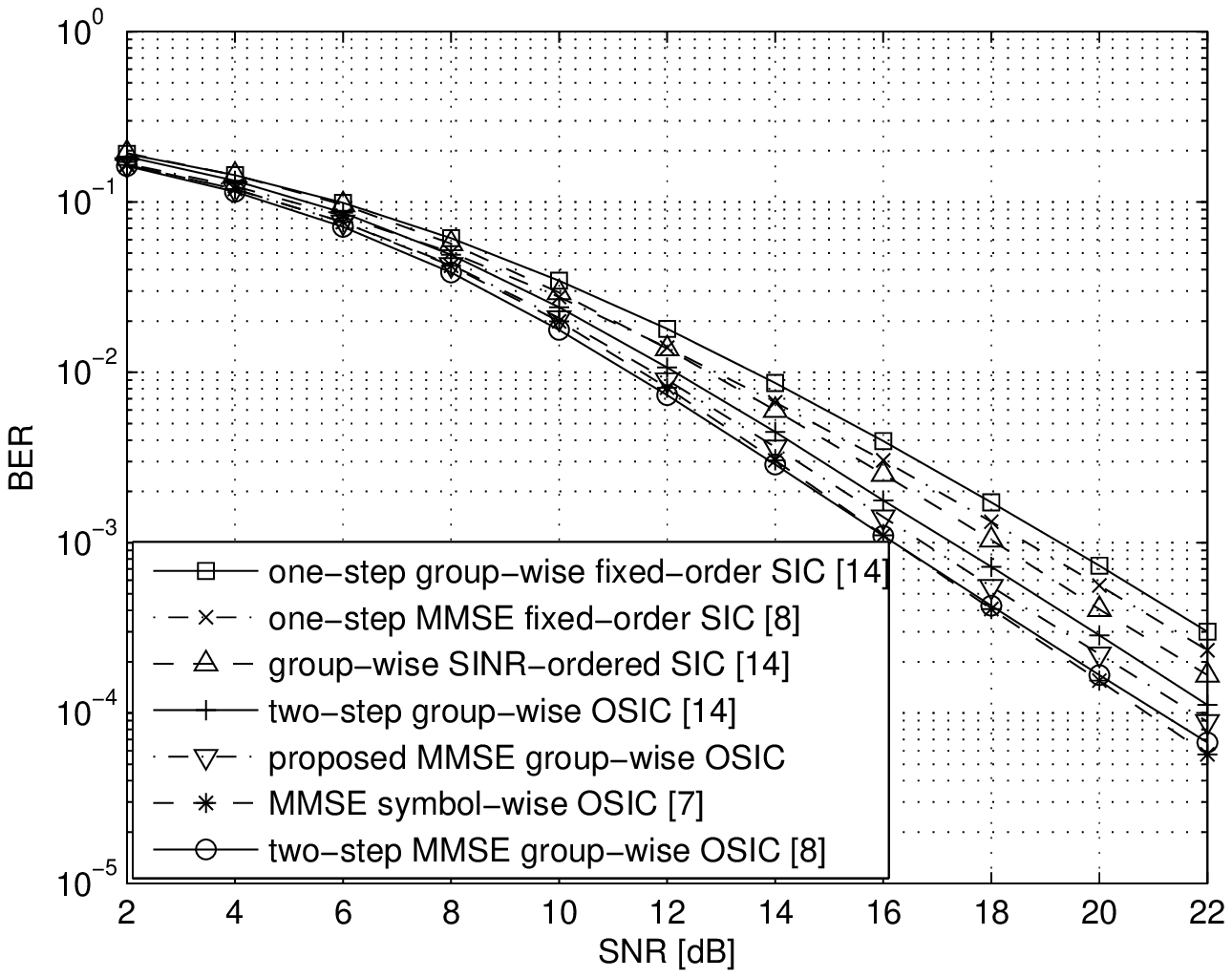}
\caption{Bit Error Rate (BER) comparison for a DSTTD system with
QPSK modulation and $2$ receive antennas.} \label{fig flops}
\end{figure}

\begin{figure}[!t]
\centering
\includegraphics[width=3.5in]{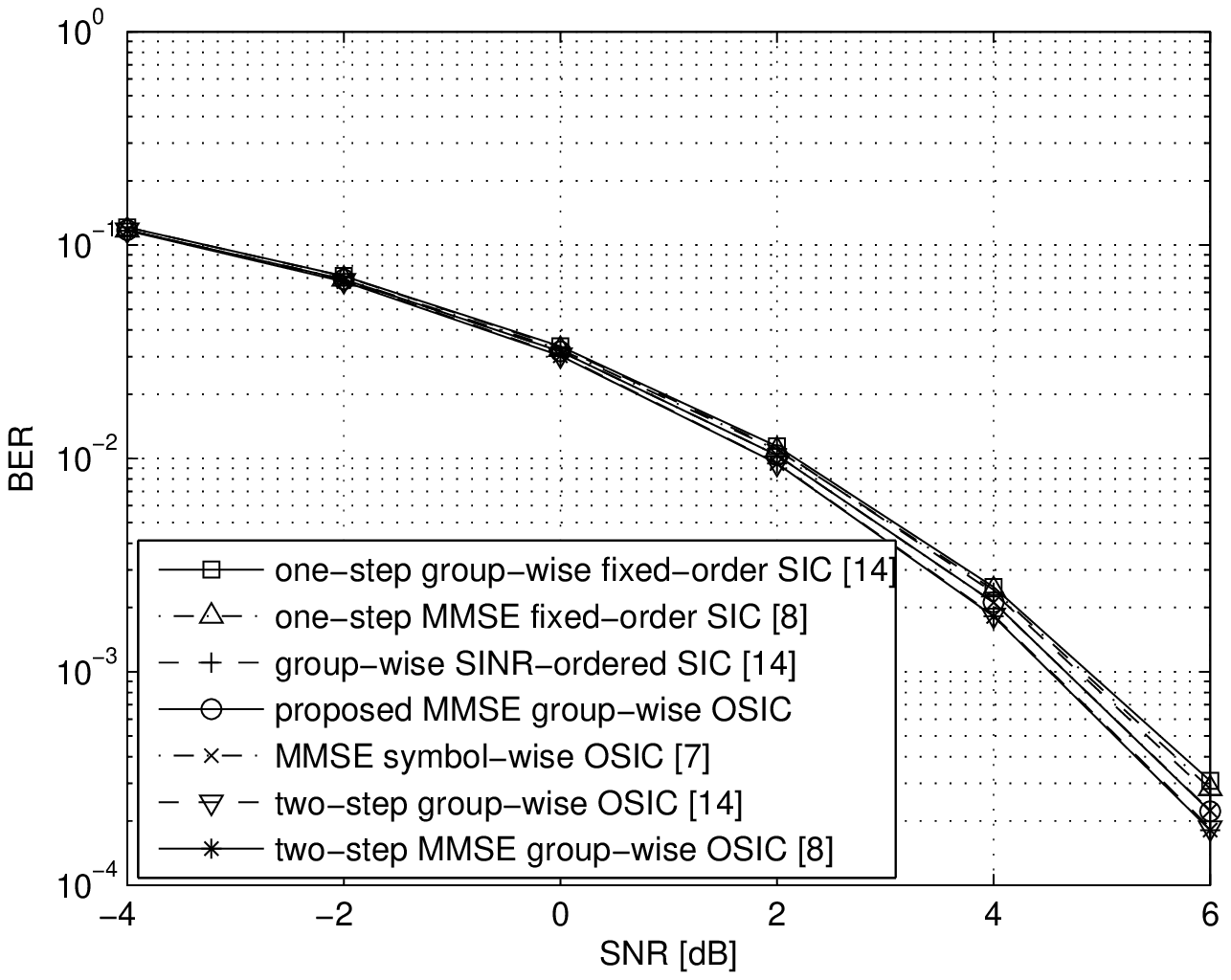}
\caption{Bit Error Rate (BER) comparison for a DSTTD system with
QPSK modulation and $8$ receive antennas.} \label{fig flops}
\end{figure}

\section{Conclusion}
We
propose a fast recursive algorithm for G-STBC, which takes full
advantage of the Alamouti structure in the equivalent channel matrix
to reduce the computational complexity dramatically.
With respect to the existing efficient algorithms for G-STBC or only
for DSTTD, the proposed group-wise MMSE OSIC algorithm for G-STBC
achieves better performance and usually requires less complexity.
When $M=N$, the proposed algorithm speeds up the MMSE sub-optimal
OSIC algorithm for G-STBC by $2.57$, speeds up the recently proposed
DSTTD algorithm by up to $4.55$, and even requires less real
multiplications than the linear ZF G-STBC algorithm.


\appendices
\section{The Derivation of (\ref{derive5})}
In what follows, ${\bm{{\rm 0}}}_{M}$ is the ${M \times M}$ zero matrix.  
Substitute (\ref{Oct21equ19a22}) into ${\bf{{\bar R}
}}_{m}{\bf{{{\bar Q}  }}}_{m}={\bf{I}}_{2m}$ to
   obtain $\left[ {\begin{array}{*{20}c}
   {{\bf{{\bar R}  }}_{m-1}^{} } & {{\bf{{\bar{v}}}}_{m-1}^{} }  \\
   {{\bf{{\bar{v}}}}_{m-1}^H } & {{{\bar {{\upsilon}}} } _{m} }  \\
\end{array}} \right]\left[ {\begin{array}{*{20}c}
   {{\bf{{{\bar T}  }}}_{m-1}^{} } & {{{\bf{\bar {w}}}}_{m-1}^{} }  \\
   {{{\bf{\bar {w}}}}_{m-1}^H } & {{{\bar {{\omega}}}} _{m} }  \\
\end{array}} \right] = {\bf{I}}_{2m}$,
from which we can deduce
\begin{subnumcases}{\label{derive4}}
{\bf{{\bar R}  }}_{{m-1}}^{} {{\bf{\bar {w}}}}_{{m-1}}  +
{\bf{{\bar{v}}}}_{{m-1}}
{{\bar {{\omega}}}} _{m}=\bm{0}_{2}, & \label{derive4a}\\
{\bf{{\bar{v}}}}_{{m-1}}^H {{\bf{\bar {w}}}}_{{m-1}}^{}  + {{\bar
{{\upsilon}}} } _{m}
{{\bar {{\omega}}}} _{m}  = {\bf{I}}_2, &  \label{derive4b}\\
{\bf{{\bar R}  }}_{{m-1}}^{} {\bf{{{\bar T}  }}}_{{m-1}}^{}  +
{\bf{{\bar{v}}}}_{{m-1}}^{} {{\bf{\bar {w}}}}_{{m-1}}^H  =
{\bf{I}}_{2(m-1)}. & \label{derive4c}
\end{subnumcases}

  On the other hand, let us extend equation (17) in \cite{QR_BLAST}
  to ${\bf{\bar {H}}}_m$ in (\ref{Jan9def_BarH_m}) for G-STBC, to obtain
\begin{equation}\label{equ:28}
{\bf{{{\bar Q}  }}}_m  = \left( {{\bf{\underline H}}_m^H
{\bf{\underline H}}_m } \right)^{ - 1} = {\bf{\bar L}}_m^{ - 1}
{\bf{\bar L}}_m^{ - H},
\end{equation}
where ${\bf{\underline H}}_m  = \left[ {\begin{array}{*{20}c}
   {{\bf{\bar {H}}}_m^T } & {\sqrt \alpha  {\bf{I}}_{2m} }  \\
\end{array}} \right]^T$ is QR decomposed into
${\bf{\bar L}}_m$ and the orthogonal $\bm{\bar \Theta} _m$, i.e.,
${\bf{\underline H}}_m  = \bm{\bar \Theta} _m {\bf{\bar L}}_m$.
${\bf{\underline H}}_m$ consists of 
Alamouti sub-blocks \cite{Invariant_Alamouti}, and so do ${\bf{\bar
L}}_m$ \cite[Lemma ``Invariance Under QR
Factorization"]{Invariant_Alamouti} and ${\bf{\bar L}}_m^{ - 1}$
\cite[Lemma ``Invariance Under Inversion"]{Invariant_Alamouti}. Then
it can be seen from (\ref{equ:28}) that ${\bf{{\bar Q}}}_{m}$
must be Hermitian with diagonal $2\times 2$
sub-blocks that are non-negative scaled multiples
  of the identity matrix and with off-diagonal $2\times 2$ sub-blocks that are  Alamouti
  matrices \cite[Property 3 in the subsection ``B. Block Matrices"]{Invariant_Alamouti}.
  Correspondingly the diagonal $2\times 2$
sub-block
${{{\bar {{\omega}}} }}_{m}$ in ${\bf{{\bar
Q}}}_{m}$ must satisfy (\ref{derive5a}).

Now we can substitute (\ref{derive5a}) and ${\bf{{{\bar Q} }}}_m =
{\bf{{\bar R} }}_m^{ - 1}$ into  (\ref{derive4a}) to obtain
(\ref{derive5b}).
 From (\ref{derive5b}) and (\ref{derive4c}), we can deduce ${\bf{{{\bar T}
}}}_{m-1}^{}  = {\bf{{{\bar Q} }}}_{m-1} - {\bf{{{\bar Q}
}}}_{m-1}^{} {\bf{{\bar{v}}}}_{m-1}^{} {{\bf{\bar {w}}}}_{m-1}^H$
and ${\bf{{{\bar Q}  }}}_{m-1}^{} {\bf{{\bar{v}}}}_{m-1}^{} = -
{{\omega}'} _{m}^{ - 1} {{\bf{\bar {w}}}}_{m-1}$, respectively, and
then the latter is substituted into the former to derive
(\ref{derive5c}). 
Moreover, substitute (\ref{derive5b}) and (\ref{equ:20}) into
(\ref{derive4b}) to deduce
$\left( {{{{\upsilon}'}}  _{m} {\bf{I}}_2 - {\bf{{\bar{v}}}}_{m-1}^H
{\bf{{{\bar Q}  }}}_{m-1}^{} {\bf{{\bar{v}}}}_{m-1}^{} }
\right){{\omega}'}_{m} = {\bf{I}}_2$,
from 
which we obtain (\ref{derive5more}).

\section{Proposed Group-wise OSIC vs.
Symbol-wise SIC with Group-wise Ordering}
 The proposed G-STBC algorithm applies (\ref{equ:27}) to compute the
 estimates of $s_{p_{m} 1}$ and $s_{p_{m} 2}$, which are the last
 two entries of
\begin{equation}\label{ApdB5}
{\bf y}_{m}={\bf \bar Q}_{|m}{\bf z}'_{m}.
\end{equation}
 In (\ref{ApdB5}), ${\bf z}'_{ m}$ satisfies \cite{My_IC_In_z}
 \begin{equation}\label{ApdB6}
 {\bf z}'_{ m}={\bf{{{\bar H}}}}_{{|m}}^H {\bf x'}^{({m})}.
  \end{equation}
 In ${\bf x'}^{({m})}$, the interferences of the $2(M-{m})$ detected symbols have been cancelled.
Then 
 \begin{equation}\label{ApdB7}
{\bf x'}^{({m})}={\bf{{{\bar H}}}}_{{|m}}{\bf s}'_{2{m}}+{\bf n},
  \end{equation}
 where the noise ${\bf n}$ is irrelevant to 
our
 discussion
 and
 can be neglected.
 Substitute (\ref{ApdB7}) into (\ref{ApdB6}), and then substitute (\ref{ApdB6})
 into (\ref{ApdB5}) to 
 obtain
${\bf y}_{m} ={\bf \bar Q}_{|m}{\bf{{{\bar H}}}}_{{| m}}^H
{\bf{{{\bar H}}}}_{{|m}}{\bf s}'_{2{ m}}={\bf \bar Q}_{|m}({\bf \bar
R}_{|m}-\alpha{\bf{I}}_{2m}){\bf s}'_{2m}$,
i.e.,
   \begin{equation}\label{ApdB8final}
{\bf y}_{m}=({\bf{I}}_{2m}-\alpha{\bf \bar Q}_{|m}){\bf s}'_{2m}.
 \end{equation}

${\bf{{\bar Q}}}_{|m}$ in (\ref{ApdB8final}) is
 with diagonal $2\times 2$ sub-blocks that are non-negative
scaled multiples of the identity matrix, as shown in Appendix A.
Thus (\ref{ApdB8final}) can be represented as
\begin{equation}\label{show_IC_notInterf}
\footnotesize {\left[ {\begin{array}{*{10}c}
   {y_{p_{1}1}}    \\
   {y_{p_{1}2}}    \\
    \vdots      \\
   {y_{p_{m}1}}   \\
   {y_{p_{m}2}} \\
\end{array}} \right]=\left[\setlength{\arraycolsep}{1.0mm}
\renewcommand{\arraystretch}{1.0}
{\begin{array}{*{10}c}
   \times & 0 &   \cdots  & \times & \times  \\
   0 & \times &   \cdots  & \times & \times  \\
    \vdots  &  \vdots &   \ddots  &  \vdots  &  \vdots   \\
   \times & \times &   \cdots  & \times & 0  \\
   \times & \times &   \cdots  & 0 & \times  \\
\end{array}} \right]\left[ {\begin{array}{*{20}c}
   {s_{p_{1}1}}    \\
   {s_{p_{1}2}}    \\
    \vdots      \\
   {s_{p_{m}1}}   \\
   {s_{p_{m}2}}   \\
\end{array}} \right]},
\end{equation}
where $\times$ denotes the non-zero entry. Now it can be seen from
(\ref{show_IC_notInterf}) that the interference
of 
$s_{p_{i} 2}$ does not affect the 
estimate of $s_{p_{i}
1}$ (i.e. ${y_{p_{i} 1}}$), and vice versa, for $i=1,2,\cdots,m$.
That is to say, in the linear MMSE estimates of ${\bf s}'_{2m}$,
each layer of STBC encoded symbols, i.e. $s_{p_{i} 1}$ and $s_{p_{i}
2}$, remain orthogonal. Thus the performance will remain unchanged
if we modify the proposed group-wise OSIC detector into the
corresponding symbol-wise SIC detector with the group-wise optimal
detection order, which detects $s_{p_{i} 1}$ immediately after the
interference of $s_{p_{i} 2}$ is cancelled.

\section*{Acknowledgment}
The authors would like to transfer their appreciation to the editor
and the reviewers for their valuable comments leading to the
improvement of this paper.

\ifCLASSOPTIONcaptionsoff
  \newpage
\fi





\begin{thebibliography}{1}

\bibitem{Jun28_MIMO}  G. J. Foschini and M. J. Gans,
``On limits of wireless communications in a fadiig environment when
using multiple antennas," \emph{Wireless Personal Communications},
vol. 6, pp.~311-335, Mar. 1998.


\bibitem{DivstMltplxTdof} L. Zheng and D. N. C. Tse, ``Diversity and multiplexing: a
fundamental tradeoff in multiple-antenna channels," \emph{IEEE
Transactions on Information Theory}, vol. 49, no. 5, pp.~1073-1096,
May 2003.


\bibitem{zhf1} P. W. Wolniansky, G. J. Foschini, G. D. Golden and R. A. Valenzuela,
``V-BLAST: an architecture for realizing very high data rates over
the rich-scattering wireless channel", \emph{URSI International
Symposium on Signals, Systems, and Electronics,} pp.~295-300, Sept.
1998.

\bibitem{fAlamoutiSTBC} S. M. Alamouti., ``A simple transmit diversity technique
for wireless communications," \emph{IEEE J. Selected Areas in
Communications}, vol. 16, no. 8, pp. 1451-1458, Oct. 1998.

\bibitem{Jun09add_GeneralSTBC} V. Tarokh, H. Jafarkhani and A. R. Calderbank, ``Space-time block
codes from orthogonal designs," \emph{IEEE Transactions on
Information Theory}, vol. 45, no. 5, pp. 1456-1467, July 1999.


\bibitem{SMSTBC1} V. Tarokh, A. Naguib, N. Seshadri and A. R. Calderbank,
``Combined array processing and space-time coding", \emph{IEEE
Transactions on Information Theory}, vol. 45, no. 4, pp. 1121-1128,
May 1999.


\bibitem{SMSTBC3}  X. Fan, H. Zhang, H. Luo and J. Huang, ``Optimal MMSE successive
interference cancellation in group-wise STBC MIMO systems",
\emph{Journal of Systems Engineering and Electronics}, vol. 17, no.
1, pp. 85-90, 2006


\bibitem{SMSTBC2} A. F. Naguib, N. Seshadri and A. R. Calderbank, ``Applications of
space-time block codes and interference suppression for high
capacity and high data rate wireless systems", \emph{IEEE Asilomar
SSC}, vol.~2, pp.~1803-1810, Nov. 1998.


\bibitem{SMSTBC4} M. Gomaa and A. Ghrayeb, ``A Low
Complexity MMSE Detector for Multiuser Layered Space-Time Coded MIMO
Systems", \emph{Canadian Conference on Electrical and Computer
Engineering}, 22-26 April, 2007.

\bibitem{SMSTBC5} A. Stamoulis, N. Al-Dhahir and A. R. Calderbank, ``Further results on
 interference cancellation and space-time block codes", \emph{IEEE Asilomar SSC}, vol.~1, pp.~257-261, Nov. 2001.


\bibitem{SMSTBC6} T. Hardtke, K. Chin and S. Sun, ``A linear groupwise STBC detector based on
Strassen algorithm", \emph{ICICS 2007}. 


\bibitem{Jun09add_bestCited14} H. Kim and H. Park, ``Efficient successive interference
cancellation algorithms for the DSTTD system," \emph{IEEE PIMRC
2005}, vol. 1, pp. 62-66, Sept. 2005.

\bibitem{Jun09add_bestCited15} F. Wang, W. Zhao and Y. Xiong, ``Ordered group interference
cancellation for quasi-orthogonal space time block codes,"
\emph{Proc. WICOM 2006}.


\bibitem{Jun09add_TransCom} S. Jung and J. Lee, ``A New ML Based Interference Cancellation Technique for Layered
Space-Time Codes", \emph{IEEE Transactions on Communications}, vol.
57, no. 4, pp. 930-936, April 2009.




\bibitem{Jun09add_wimax} \emph{IEEE Std 802.16e-2005}, December
2005.



\bibitem{WXD_revier1_ref} N. Prasad, M.K. Varanasi, L. Venturino, X. Wang, ``An
Analysis of the MIMO SDMA Channel With Space Time Orthogonal and
Quasi-Orthogonal User Transmissions and Efficient Successive
Cancellation Decoders", \emph{IEEE Transactions on Information
Theory}, vol 54, no. 12, pp. 5427-5446, Dec. 2008.

\bibitem{QR_BLAST} R. Buhnke, D. Wubben, V. Kuhn and K. D. Kameyer, ``MMSE
extension of V-BLAST based on sorted QR decomposition", \emph{IEEE
Vehicular Technology Conference 2003-Fall}.











\bibitem{Invariant_Alamouti} A. H. Sayed, W. M., Younis and A. Tarighat, ``An invariant matrix structure in multiantenna
communications", \emph{IEEE Signal Processing Letters,} vol. 12, pp.
749-752, Nov. 2005.


\bibitem{My_IC_In_z} H. Zhu, Z. Lei, F. P. S. Chin, ``An improved recursive algorithm for
BLAST", \emph{Signal Processing}, vol. 87, no. 6, pp. 1408-1411,
Jun. 2007.


\bibitem{zhf6} Y. Shang and X. G. Xia, ``On Fast Recursive Algorithms For V-BLAST With Optimal Ordered SIC Detection",
 \emph{IEEE Transactions on Wireless Communications}, vol. 8, pp. 2860-2865, June 2009.


\bibitem{TaiwanPaper} T. Liu and Y. Liu, ``Modified fast recursive algorithm for efficient MMSE-SIC detection
of the V-BLAST system", \emph{IEEE Transactions on Wireless
Communications}, vol. 7, pp. 3713-3717, Oct. 2008.

\bibitem{Reviewer2_Journ_real_Gstbc} C. L. Ho, J. Y. Wu, T. S. Lee,
 ``Group-wise V-BLAST detection in multiuser space-time dual-signalling wireless systems", \emph{IEEE Trans.
Wireless Commun}, vol. 5, no. 7, pp. 1896-1909, July 2006.

\bibitem{Reviewer2_ML_receiver} J. Cha, S. Hur, C. Min, H. Lee, J.
Kang, ``Efficient Modified Fano Detection with Reduced Branches for
DSTTD System", \emph{IEEE International Conference on Communications
2006}.

\bibitem{Our_comments} H. Zhu and W. Chen, ``Comments on a New ML Based Interference Cancellation Technique for Layered
Space-Time Codes," \emph{IEEE Transactions on Communications}, vol.
58, no. 11, pp. 3054-3055, Nov. 2010.





\end{thebibliography}
\end{document}